\input harvmac
\pretolerance=10000
\def\ket#1{| #1 \rangle}

\Title{HWS-9719, hep-th/9710002}
{\vbox{\centerline{Duality, Partial Supersymmetry, and}
       \vskip2pt\centerline{Arithmetic Number Theory}}}

\centerline{Donald Spector\footnote{$^\dagger$}
{spector@hws.edu}}
\centerline{Department of Physics, Eaton Hall}
\centerline{Hobart and William Smith Colleges}
\centerline{Geneva, NY 14456 USA}

\vskip .3in
We find examples of duality among quantum theories that are related to
arithmetic functions by identifying distinct Hamiltonians that have
identical partition functions at
suitably related coupling constants or temperatures.
We are led to this after first developing the
notion of partial supersymmetry---in which some, but not all, of the
operators
of a theory have superpartners---and using it to construct fermionic and
parafermionic thermal partition functions, 
and to derive some number theoretic
identities. In the process, we also find a bosonic analogue of the
Witten index,
and use this, too, to obtain some number theoretic results related to
the
Riemann zeta function.
\vskip .5in
\centerline{PACS 02.10.Lh, 05.30.-d, 11.30.Pb, 11.25.-w}

\Date{8/97; rev. 11/97}

\newsec{Introduction}

Arithmetic quantum theories --- theories for which the partition
function
is related to the Riemann zeta function or other Dirichlet
series --- have yielded
compelling results in mathematical 
physics \ref\numqm{ 
    B. Julia, {\it J. Phys. France }{\bf 50} (1989) 1371\semi
    D. Spector, {\it Phys. Lett. }{\bf A 140} (1989) 311\semi
    B. Julia, ``Statistical Theory of Numbers,'' in {\it Number
Theory and Physics}, edited by J.M. Luck, P. Moussa, and M. Waldschmidt,
{\it Proceedings in Physics }{\bf 47} (Springer-Verlag, Berlin, 1990)
276\semi
    D. Spector, {\it Commun. Math. Phys. }{\bf 127} (1990) 239.
}\ref\bowick{
I. Bakas and M. Bowick,  {\it J. Math. Phys.} {\bf 32}(1991) 1881\semi
J.-B. Bost and A. Connes, {\it Selecta Mathematica} {\bf 1} (1995)
411\semi
A. Connes, 
{\it C. R. Acad. Sci. Paris, Ser. I Math.} {\bf 323} (1996) 1231\semi
P. Contucci and A. Knauf, {\it J. Math. Phys.} {\bf 37} (1996) 5458.
}.
In addition to providing physical derivations and
interpretations of various number theoretic results, such as an
understanding of the M\"obius inversion formula based on supersymmetry,
these theories possess, as does string theory, a Hagedorn temperature,
and so
these theories may serve as toy models for certain aspects of
string theory.

In this paper, I find further evidence of the usefulness of arithmetic
quantum thories in these areas.  Most importantly, 
I find examples of duality in
arithmetic theories, that is, of distinct arithmetic theories
at different couplings that in fact have the same partition function.
Such dualities are analogous to the
coupling constant dualities of string theory.\foot{See Section VII for a 
fuller discussion of this comparison.} In
addition, I find physical underpinnings for additional number theoretic
results.  The most interesting examples here rest on ``partial
supersymmetry,''
a property identified in this paper, 
in which some, but not all, of the creation
and annihilation operators have superpartners.  This leads not only to 
a physical basis for some number
theoretic results, but also to a formulation of parafermions of
non-integer
order and to the identification of an analogue for the Witten index
based on a
bosonic grading.

\newsec{Review of Notation and Ideas}

We here review the notation and way of thinking that allow a connection
to be drawn between arithmetic number theory and quantum field theory.

Consider some non-interacting quantum field theories defined
in terms of their creation and annihilation operators.  Let
$b^\dagger_k$
and $b_k$ denote bosonic creation and annihilation operators, and
let $f^\dagger_k$ and $f_k$ denote fermionic creation and annihilation
operators.  In addition, we label the prime numbers $p_k$ in order
of increasing magnitude, so that $p_1=2$, $p_2=3$, $p_3=5$, etc.

For use throughout this paper, we define the bosonic Hamiltonian
\eqn\hbdef{
H_B = \omega \sum_{k=1}^\infty \log(p_k) b^\dagger_k b_k~~~,}
and the fermionic Hamiltonian
\eqn\hfdef{
H_F = \omega \sum_{k=1}^\infty \log(p_k) f^\dagger_k f_k~~~.}
We label the states of such theories using a G\"odel numbering.  Let
us denote the vacuum state $\ket{1}$. 
For the bosonic theory, we label the states\foot{For
simplicity, we use the atypical convention that the creation
and annihilation operators produce normalized states.}
\eqn\godelket{
(b^\dagger_{i_1})^{r_1}(b^\dagger_{i_2})^{r_2}\cdots
     (b^\dagger_{i_n})^{r_n}\ket{1} = \ket{N}~~~,}
where $N=(p_1)^{r_1}(p_2)^{r_2}\cdots (p_n)^{r_n}$.
We may use a similar labeling for the states of the fermionic theory.  
Note that
with this labeling, the Fock space of states for the bosonic field
theory is labeled by the positive integers, while
the Fock space of states for the fermionic field theory is 
labeled by the squarefree positive integers.
One sees \numqm\ that
\eqn\oldbose{
\tr e^{-\beta H_B}=
\sum_{m=1}^\infty{1\over m^{\beta\omega}}=\zeta(\beta\omega)}
and
\eqn\oldfermi{
\tr (-1)^F e^{-\beta H_F}=\sum_{m=1}^\infty{\mu(m)\over
m^{\beta\omega}}~~~,}
where $\mu(m)$ is the M\"obius inversion function \ref\mobidef{
T.M. Apostol, {\it An Introduction to Analytic Number Theory} (New York:
Springer-Verlag 1976)\semi see also M. Aignier, {\it Combinatorial
Theory}
(New York: Springer-Verlag 1979).}.
Supersymmetry tells us, via the Witten 
index \ref\windex{E. Witten, {\it Nucl. Phys.}  {\bf B202} (1982) 253.},
that $\tr \bigl[(-1)^F e^{-\beta(H_B+H_F)}\bigr]=1$, and so
\eqn\muzeta
{\sum_{m=1}^\infty{\mu(m)\over m^{\beta\omega}}
    =\tr [(-1)^F e^{-\beta H_F}]={1\over \zeta(\beta\omega)}~~~.}

\newsec{Thermal Partition Functions}

In the preceding work, we have identified the bosonic thermal partition 
function with the Riemann zeta function, but only the graded fermionic
partition function with a number theoretic function.  On the other hand, 
the physics of a theory is governed by the genuine thermal partition 
function, and so to understand the thermodynamics
of the fermionic field theory of
$H_F$,  we need to evaluate the partition function
\eqn\partf
{Z_F(\beta,\omega)=\tr ~ e^{-\beta H_F}~~~.}
To calculate this partition function, we will introduce a notion of
partial supersymmetry. This will then give us a physical and algebraic
understanding for a particular zeta function identity.\foot{This
partition
function, like the parafermionic one of the next section,  was obtained
through
number theoretic arguments in \bowick.  Thus it is our method in this
section
that is new, not the result.}

Our approach is to start with the bosonic partition function and
to cancel out just the right states so that we are left with the
fermionic
partition function.
Suppose, then, that we define operators $q_k = (b_k)^2$ and
$q^\dagger_k=
(b^\dagger_k)^2$, and in addition we define operators $c_k$ and
$c^\dagger_k$ which have the same effect as $b_k$ and $b^\dagger_k$
(respectively), except that $c_k$ and $c^\dagger_k$ both square to zero.
It is easy to see, then, that we can write the bosonic Hamiltonian
\eqn\althb
{H_B = \omega \sum_k \log(p_k) c^\dagger_k c_k
      +\omega \sum_k 2\log(p_k) q^\dagger_k q_k~~~.}
The two portions of this decomposition of $H_B$ commute, and so
\eqn\prodone
{\tr e^{-\beta H_B} = \tr [e^{-\beta \omega \sum_k \log(p_k) c^\dagger_k
c_k}]
                    ~\tr [e^{-\beta \omega \sum_k 2\log(p_k) q^\dagger_k
q_k}]}
Because the $c_k$ and $c^\dagger_k$ square to zero, the first term in
this
product is exactly the fermionic partition function.

Notice that
the $q^\dagger_k$ and $q_k$ operators just appear as ordinary creation 
and annihilation operators, albeit associated with energy $2\log(p_k)$. 
(Due to
our non-conventional normalization convention,
there is no problem with how the resulting states are normalized.)
Adding to the original Hamiltonian the expression
$\omega\sum_{k=1}^{\infty}2\log(p_k)f^\dagger_k f_k$, we are in a
position
to cancel out all but the $c_k$ and $c^\dagger_k$ terms by taking a
graded 
sum, by means of a Witten
index type of result. It is this situation, in which we have fermionic
superpartners for some of the bosonic creation and annihilation
operators
(the $q^\dagger_k$ and $q_k$) but not others (the $c^\dagger_k$ and
$c_k$)
that we term {\it partial supersymmetry}.  We thus find that 
\eqn\getting{\eqalign{
  Z_F(\beta,\omega) &=
  \tr [e^{-\beta \omega \sum_k \log(p_k) c^\dagger_k c_k}]\cr
   &=\tr[(-1)^Fe^{-\beta \omega \sum_k 
  ( \log(p_k) c^\dagger_k c_k + 2\log(p_k)q^\dagger_k q_k
       + 2\log(p_k)f^\dagger_k f_k)}]\cr
   &=\tr[(-1)^Fe^{-\beta(H_B+2\beta H_F)}]~~~.}} 
The bosonic and fermionic pieces in the final expression
in \getting\ factorize into
pieces that are easy to evaluate (see \numqm), leading to the number
theoretic result \eqn\ntres
{Z_F(\beta,\omega)
     =\sum_{m=1}^\infty {|\mu(m)|\over m^s} = {\zeta(s)\over
\zeta(2s)}~~~,} where $s=\beta\omega$.
Notice that we have obtained this number theoretic identity using a
partial supersymmetry of a quantum
mechanical theory, rather than the conventional explicit calculations
of number theory using infinite products.

Our use of partial supersymmetry to reduce a bosonic to a fermionic
theory appears calculationally
much like a BRST 
\ref\brstref{C. Becchi, A. Rouet, and R. Stora, {\it Phys. Lett. }{\bf
52B}
  (1974) 344\semi
I.V. Tyupin, Lebedev preprint FIAN No. 39 (1975)\semi
 C. Becchi, A. Rouet, and R. Stora, {\it Ann. Phys. }{\bf 98} (1976)
287.}
cancellation.  However, there is no
gauge symmetry that is enforcing these cancellations, and so we stick
to the term ``partial supersymmetry.''
If one could identify a gauge invariance
underlying this result, then one could perhaps recast these results as
proper
BRST cancellations.  Such a result would undoubtedly provide further
deep insights into the structure of these number theoretic quantum
models.

\newsec{Parafermions of Arbitrary Order}

A natural generalization of the above results is to consider
\eqn\requation
{Z_r(\beta,\omega)=\tr [(-1)^F e^{-\beta(H_B+rH_F)}]~~~.}
As we know, $Z_1=1$, as the $H_F$ terms cancel all the $H_B$ terms,
yielding
the Witten index of theory, while, as we just saw,
$Z_2=Z_F$, because the
fermionic terms cancel all quadratic and higher powers 
of the original bosonic creation operators.

For higher integers $r$, we can identify a partial supersymmetry as
well.
Let us define $\chi^\dagger_k$ and $\chi_k$ such that their $r^{th}$
powers,
but no lower powers, vanish as operators, and let
$r^\dagger_k=(b^\dagger_k)^r$ and $r_k = b_k^r$.  Thus 
\eqn\hbwithr{
H_B=\omega\sum_{k=1}^\infty\bigl({{\log(p_k)\chi^\dagger_k\chi_k
                    +\log(p_k^r)r^\dagger_k r_k}\bigr)~~~.}}
Then the fermionic terms in the expression for $Z_r$ \requation\ cancel
the $r^\dagger_k$ and $r_k$ terms in $H_B$, and so
\eqn\paraz
{Z_r(\beta,\omega)
   =\tr [(-1)^F e^{-\beta(H_B+rH_F)}]=\tr e^{-\beta H_r}~~~,}
where $H_r = \omega\sum_{k=1}^\infty\log(p_k)\chi^\dagger_k\chi_k$ is
the
Hamiltonian for order $r$ parafermions for the logarithmic spectrum we
are studying.  We can evaluate this expression, as the bosonic and
fermionic pieces of \requation\ are non-interacting and hence factorize.
One sees that the partition function for parafermions of order $r$,
i.e.,
objects that are subject to an exclusion principle that states that no
more than $r-1$ parafermions can have the same quantum numbers,  is
given
by
\eqn\zrvalue
{Z_r(\beta,\omega)={\zeta(\beta\omega)\over\zeta(r\beta\omega)}~~~.}
What is compelling about this result is that we have found it using
partial supersymmetry.

One knows, of course, that the bosons are the infinite order
limit of parafermions.  It is interesting to see how that appears in
this
context.  As we take $r$ to infinity, the energies of states with
fermionic
excitations go to infinity, and so the fermionic creation and
annihilation
operators decouple \ref\decoupling{T. Appelquist and J. Carazone, 
{\it Phys. Rev. }{\bf D11} (1975) 2856.},
leaving behind the purely bosonic partition function.  Thus the recovery
of the bosonic theory at infinite $r$ is, in our language, a decoupling
effect.

One of the most intriguing features of the partition function $Z_r$ is
that it is well-defined for non-integer values of $r$.  We do not simply
mean that non-integer values of $r$ can be plugged into the zeta
function
expression \zrvalue.  Rather, we can put non-integer values of $r$
into \requation\ quite sensibly and naturally---after all, why should 
the parameters of a Hamiltonian be integral---and this yields, from an
expression with ordinary bosons and fermions, a partition function for
exotic parafermions, parafermions of non-integral order.
Thus, although there is no simple exclusion principle that one could
state for such exotic parafermions, nor a simple Hamiltonian directly
in terms of creation and annihilation operators for such exotic
parafermions,
we can obtain the thermal partition function quite easily from
\eqn\arbz
{Z_r(\beta,\omega)=\tr \bigl[(-1)^F e^{-\beta (H_B+rH_F)}\bigr]~~~.}

What is the meaning of such theories?  On the one hand, the partition 
function tells us what the theory is, so any quest for further 
interpretations might be viewed as superfluous.  On the other hand,
understanding what non-integral parafermions mean in
a practical sense is physically intriguing.  
It is worth observing that
the ability to vary the parameter $r$ continuously in \paraz\ means that
we have a way to vary continuously from conventional parafermions of one
integer order to another.  This should enable one to connect results
in the different conventional parafermion theories, presumably through
analyticity or topological arguments.

Note that for parafermions of order $r$ real and less than $1$,
there is a temperature less than the Hagedorn
temperature \numqm 
\ref\hagedorn{R. Hagedorn, {\it Nuovo Cim.}  {\bf 56A} (1968) 1027.}
where the thermal partition function vanishes, indicating that
the physical sense of these theories is limited to $r>1$.
This is no surprise.  At $r=1$, the corresponding order $1$ parafermions
obey
an exclusion principle
such that states with one or more excitations are forbidden,
which means that only the vacuum state is physical. (Indeed, $r=1$
is simply the case of the
Witten index calculation, in which all states but the vacuum state
are canceled out.)  Phrased in this way, parafermions of order $r<1$
seem
clearly to be physically untenable, as well as mathematically
troublesome. 

\newsec{Supersymmetry and Boson Number}

Recall from \numqm\ that the mathematical interpretation
of the supersymmetric Witten index result that
\eqn\zetamu
{\tr e^{-\beta H_B}~\tr [(-1)^F e^{-\beta H_F}] =1~~~}
is 
\eqn\mathmz
{\sum_{m=1}^\infty {\mu(m)\over m^s}={1\over \zeta(s)}~~~,}
while we also know from \zetamu 
that the graded fermionic partition function is the inverse of
the thermal bosonic partition function.
Now that we have just calculated the thermal fermionic partition
function, 
we can ask what its inverse is.  We will find the result physically,
and give its corresponding mathematical interpretation.

Supersymmetry, we know, pairs bosonic and fermionic states.  Consider
the Hamiltonian $H_S = H_B+H_F$.
Supersymmetry pairs each $b_k$ with $f_k$ and each $b^\dagger_k$ with
$f^\dagger_k$. Let us define $(-1)^{N_B}$ to be the operator that has
value $+1$
for Fock space states with an even number of bosonic creation operators,
and
$-1$ for Fock space states with an odd number of bosonic creation 
operators. 
Note that $(-1)^{N_B}$ commutes with $f^\dagger_k$ and $f_k$.  Because
the
supercharges of $H_S$
involve terms like $b^\dagger_k f_k$ and  $f^\dagger_k b_k$,
all states with non-zero energy come in
superpartnerships with an equal number of even and odd $N_B$ states.
Only the zero energy states need not be paired in this way, as they
are the supersymmetry singlet states. 
Consequently, for the theory with Hamiltonian $H_S = H_B+H_F$. we see
\eqn\altind {\tr [(-1)^{N_B} e^{-\beta(H_B+H_F)}] =1~~~,}
which is analogous to but distinct from the familiar Witten index result 
for  $\tr [(-1)^F e^{-\beta H_S}]$.
Since $H_B$ and $H_F$ commute with each other,
we see that
\eqn\almost{\tr e^{-\beta H_F}={1\over\tr [(-1)^{N_B} e^{-\beta
H_B}]}~~~.}
The inverse of the thermal fermionic partition function is the graded
bosonic partition function.

The mathematical translation of this is as follows.  Using the G\"odel
numbering described above, we can represent $(-1)^{N_B}$ by the
arithmetic function $\psi(m)$, where $\psi(m)$ is totally multiplicative
(i.e., $\psi(m)\psi(n)=\psi(mn)$ for all natural numbers $m$ and $n$)
and
with $\psi(p^r)=(-1)^r$ for any prime number $p$.  Thus, using the
connection with Dirichlet convolution describe in \numqm, we see that
\eqn\psimu
{\psi*|\mu| (m) = {\bf 1}(m),}
where {\bf 1} is the identity function for Dirichlet convolution,
${\bf 1}(m) = \delta_{m,0}$, or, equivalently,
\eqn\psimueq
{\sum_{m=1}^\infty {\psi(m)\over m^s} =
 {1\over\sum_{m=1}^\infty {|\mu|(m)\over m^s} }
  = {\zeta(2s)\over\zeta(s)~~~.} }

Thus we have used our thinking about number theory to recognize that,
generically, we can introduce a bosonic grading quite analogous to the
fermionic grading, and this has led us physically to another number
theoretic identity, namely \psimueq.  We note
that the bosonically graded index is topological in nature, unaffected
by
deformations of the parameters of a theory as long as long as the
supercharges
remain odd under $(-1)^{N_B}$, and thus it is reliably calculable even
by
approximation methods.  Supersymmetry is unbroken if this bosonic index
is
non-zero. Note, too,
that $(-1)^F$ corresponds to the arithmetic function $\mu$, and
$(-1)^{N_B}$ to the arithmetic function $\psi$.  One learns from
supersymmetry
that the inverses of these functions under Dirichlet convolution obey
a nice symmetry: $\mu^{-1}=|\psi|$, while $\psi^{-1}=|\mu|$.  It would
be interesting to find other pairs of functions with this type of 
partnering when taking the inverse, and to see if there is a
supersymmetric
reason underlying such a relationship.

\newsec{Graded Parafermion Partition Functions}

Our discussion of graded partition functions leads us naturally
to consider graded parafermion partition functions.  Here we present
some
results, but leave the detailed calculation to the readers, and reserve
speculation on these results for elsewhere.

Consider the parafermions of order $r$ discussed previously.
Define the operator $\hat\eta$ such that a one-parafermion state has
$\hat\eta$-eigenvalue $\eta$, where $\eta^r=1$. 
Then
\eqn\here{
\tr [\hat\eta e^{-\beta H_r}] 
    = \tr [\hat\eta  e^{-\beta H_B}]~\tr [(-1)^Fe^{-\beta rH_F)}]~~~,}
using our representation of parafermions \paraz\ and 
partial supersymmetry.  Note that the
bosonic creation operators also have $\hat\eta$-eigenvalue $\eta$,
while the fermionic creation operators are $\hat\eta$-neutral.

There is an alternative graded partition function that one can
naturally define.  
Let $\hat\xi$ be an operator such that a one-parafermion
state has $\hat\xi$-eigenvalue $\xi$ where $\xi^r=-1$.
Then using our method of partial supersymmetry and the notion
of the bosonic index \altind, we find that
\eqn\there{
\tr [\hat\xi e^{-\beta H_r}]
    = \tr [\hat\xi e^{-\beta H_B}]~\tr [e^{-\beta r H_F}]~~~.}
The convention, again, is that the bosonic creation operators have
$\hat\xi$-eigenvalue $\xi$, while the fermionic creation operators
are $\hat\xi$-neutral.

The $\hat\eta$- and $\hat\xi$-graded partition functions 
can be written in these factorized forms involving products of purely
bosonic and purely fermionic terms because of
existence of a partial supersymmetry in these expressions, which leads
to cancellations along the lines of the Witten index in \here\ and
of the bosonically graded index \altind\ in \there.  Other gradings
would not admit such a factorization.

\newsec{Duality}

The identities we have obtained above provide relationships for
partition functions (thermal and graded) at different temperatures.
This offers an interesting possibility, akin to duality in 
supersymmetric and string theories: that we can identify two distinct
theories at distinct temperatures that nonetheless give rise to
the same thermodynamics.  We will in fact realize this possibility, with
an
example in which the Euler totient function plays a central role.

Before proceeding further, it is worth commenting on our use of the
term ``duality.''  In string theory, this term is most conventionally
used to refer to relationships that directly rate the weak and strong
coupling regimes (roughly speaking, replacing a coupling constant $g$ by
$1/g$).  Here, we use the term in a somewhat broader sense, to refer
to identities among theories at different coupling constants in general.
Although the relationships we identify do not simply involve inverting a
coupling constant, they can be iterated to related weak and strong
couplings.
Furthermore, even without such iteration, the identities we find between
arithmetic quantum theories can clearly play a role analogous to the
role played by weak-strong coupling dualities in string theory,
establishing the equivalence of superficially distinct theories, and
thus
allowing the use of one formulation of a theory to analyze  another.
Hence we employ the term ``duality'' to characterize the identities
among
arithmetic quantum theories that we establish below.

To begin our analysis,
then, consider the thermal partition functions associated
with the original Hamiltonians $H_B$ and $H_F$, which we will
denote as $Z_B$ and $Z_F$,
respectively.  Then \ntres\ can be interpreted as
\eqn\twotemp{
Z_F(\beta) Z_B(2\beta) = Z_B(\beta)~~~.}
Alternatively, if we denote the fermionic partition function graded
with $(-1)^F$ as $\Delta_F$, we have
\eqn\delver{\Delta_F(2\beta)=Z_F(\beta)\Delta_F(\beta)~~~.}
What is interesting about these expressions is that they
relate partition functions at different temperatures.
There are two interpretations possible here:

(a) In \twotemp, we have a non-interacting mixture of two systems, each
at its own temperature, equivalent to a third system at its own 
temperature;

(b) In \delver, we have a mixture of two systems thermalized with each
other
at one temperature equivalent to a different system at a different 
temperature.

Note that in (b), not all the partition functions are thermal (some are
fermionically
graded), while in (a) we have an unthermalized mixture of two systems.
Ideally, we would like a thermalized system at one temperature
equivalent
to a thermalized system at a different temperature.  It turns out
that this is possible to achieve.

Consider the Euler totient function,
defined by $\varphi(m)=\sum_{d: (d,m)=1}1$, i.e., it
is the number of positive integers less than $m$ that are relatively
prime
to $m$.  Then it is a standard result in number theory
that
\eqn\dirtot
{{\zeta(s-1)}=\zeta(s)\sum_{m=1}^\infty {\varphi(m)\over m^s}~~~.}
We can interpret this physically as follows.  Let us imagine two
theories: one is governed by the original Hamiltonian $H_B$ 
from \hbdef\ above.
The other has the same spectrum as $H_B$, but has degeneracies given by
the totient function; it has $\varphi(m)$ states with energy
$\omega\log(m)$. 
We call the Hamiltonian for this theory $H_\varphi$ and its partition
function
$Z_\varphi$.

Then we see that, in terms of the dimensionless 
parameter $s=\beta\omega$,
\eqn\phystot{
Z_B(s-1)= Z_B(s) Z_\varphi(s)~~~.}
In other words, a thermalized mixture of two ideal gase, governed by 
the Hamiltonians $H_B$ and $H_\varphi$, is equivalent simply to
a higher temperature gas governed by $H_B$.
Alternatively, we can phrase this as a coupling constant duality.  The
Hamiltonian $H_B$ at coupling constant $\omega-{1\over\beta}$ and the
Hamiltonian $H_\varphi$ at coupling constant $g$ have the same partition
function at temperature $T=\beta^{-1}$.  The presence of this duality
indicates an important additional similarity connecting arithmetic
quantum
theories and string theories.

The duality described above is clearly intriguing.  
There are two especially interesting directions to consider.

(1) These expressions give a way to shift temperature via coupling new
systems.  Let us call a system governed by Hamiltonian $H_B$ a Riemann
gas, and one governed by $H_\phi$ a totient gas.  Then what we have
found
is that if we start with a Riemann gas at one temperature, and then mix
it with a totient gas at the same temperature, that this is equivalent
simply to taking the Riemann gas and increasing its temperature a
certain
amount.  Note that we can iterate this process by mixing in additional
totient gases, and thereby shift the temperature further.
This gives us an alternative to standard methods such as
the renormalization group for thinking about how to shift temperatures;
relating such distinct methods is potentially powerful.  The most
interesting outcome would be to determine a way to generalize this 
notion of obtaining a temperature shift by mixing in a new quantum
system,
finding ways to apply it to theories not so closely tied 
to the Riemann zeta function.

(2) We have now identified, among the class of arithmetically-based
quantum models, theories that exhibit duality behavior in their
partition
functions, with the best example being the duality \phystot\
occurring for the theory with
Hamiltonian $H_B$.  We also know that this same theory exhibits a
Hagedorn 
temperature.  Both these traits---existence of duality and existence of
a Hagedorn temperature---are distinguishing characteristics of string
theory.
This adds further support to the notion that the $H_B$ theory and its
relatives
are related in a  meaningful way to string theory, and thus that 
exploration of these arithmetic theories
will prepare us to understand better the nature of string theory.

We can expand on this idea a bit further.  It is well-known that
in string theory, the duality between large and small radius effectively
means that there is a minimum distance in string theory.  On the other
hand,
the Hagedorn temperature means there is a maximum temperature in a
theory.
Perhaps, in some similar fashion, the existence of a maximum temperature
is associated with a temperature duality (recall that the duality we
found
could be understood as a temperature or as a coupling constant duality)
in and of
itself.  These number theoretic field theories offer us an arena other
than
string theory in  which to explore this question.

\newsec{More with Partial Supersymmetry}

In earlier sections, we found that we could evaluate certain physically
interesting partition functions, and simultaneously obtain some
interesting
number theoretic identities, using partial supersymmetry, in which some
but not
all of the bosonic creation operators had fermionic superpartners.
There are other
number theoretic results which appear as if they should have an
interpretation
via partial supersymmetry.  We give here two particular examples to show
the
kinds of results that it appears ought to be possible to derive via
partial
supersymmetry, although to date no such derivation exists.

As a first example, consider the function
\eqn\defsig{\sigma_\alpha(m)=\sum_{d|m}d^\alpha~~~,}
where the summation is over all positive integers $d$ that are factors
of the positive integer $m$.
The Dirichlet series of this function is related to the Riemann zeta
function
by
\eqn\dirsig{
\zeta(s-\alpha)
={1\over\zeta(s)}\sum_{m=1}^\infty{\sigma_\alpha(m)\over m^s}~~~.}
Let us define $H_{\sigma_\alpha}$ to be the Hamiltonian that has as
its spectrum $\sigma_\alpha(m)$ states of energy $\omega\log(m)$.
Then\eqn\sigcan{
\tr [e^{-(\beta-{\alpha\over\omega})H_B}]
      =\tr [(-1)^F e^{-\beta(H_{\sigma_\alpha}+H_F)}]~~~.} 
The right side involves a partial supersymmetric cancellation (note that
none of the degeneracies go to zero) at temperature $\beta^{-1}$,
resulting
in a theory a different temperature.

A second representative case comes from
re-writing the identity \dirtot\ of the preceding section as
\eqn\rephi{
Z_\varphi(\beta,\omega)={\zeta(s-1)\over\zeta(s)}
                       =\tr\bigl[ e^{-(\beta-{1\over\omega})H_B}\bigr]
                         ~\tr\bigl[ (-1)^F e^{-\beta H_F}\bigr]~~~.
}
In this context, the Euler totient function appears as the
degeneracy factor resulting after a cancellation between bosons and
fermions at
different temperatures.  This is a variation on the
partial supersymmetry used above; here,
if the bosons and fermions were at the same temperature, the
cancellation would
be complete. Mathematically, the right side of \rephi\ resembles a
partition
function for a supersymmetric Hamiltonian with non-supersymmetric
boundary
conditions.

Needless to say, these two examples could both be written as identities
involving a single temperature but different coupling constants.

The importance of highlighting these identities is to demonstrate what
partial
supersymmetry can offer.  If these and a host of similar identities
involving ratios of zeta functions and other Dirichlet series can be
understood
in this way, partial supersymmetry promises to be a valuable new tool
for enriching our
understanding of arithmetic number theory.

\newsec{Conclusions}

Arithmetic quantum theories have provided some of the most aesthetically
pleasing examples of connections between mathematics and physics, and in
this
paper, we have found that these connections are even more compelling and
worthy
of study.  

Most importantly, we have found dualities among the partition functions
for
arithmetic quantum theories, dualities that can be interpreted as
coupling
constant or temperature dualities, demonstrating the equivalence of
apparently
distinct Hamiltonians.  Our work has suggested that there may actually
be
reason to expect a connection between some types of dualities and the
existence
of a Hagedorn temperature, an area that warrants further investigation. 
With
these dualities, it is possible to shift the temperature of a certain
theory
by mixing in additional quantum excitations at the original temperature;
relating this interpretation to more familiar methods such as the
renormalization groups should also 
provide useful insights into or constraints on
the behavior of these theories. We are interested, too, in finding more
useful
applications of the bosonically graded topological index.

On the technical side, the most significant result of this paper is the
identification of the technique of partial supersymmetry.  This has
given
a physical basis for a variety of number theoretic results, and we have
identified further types of results that should yield to such an
analysis. 
Indeed, one would expect this method to be useful beyond the realm of
arithmetic
quantum theories.

This technique of partial supersymmetry has given us a physically-based
formalism for interpolating from one Dirichlet series to another using
ordinary
bosons and fermions.  Indeed, it is this idea that underlies our
formulation of
non-integer order parafermions. This is a potentially powerful idea, as
it may
allow the use of topological insights to study
arithmetic number theory, by using continuity  arguments to connect the
behavior
of different Dirichlet series.

It is our hope that the ideas introduced in this paper will be of value
not
only in understanding arithmetic quantum theories and directly related
examples
(physical ones such as string theory, mathematical ones such as
Dirichlet
series), but may also be extended to wider range of theories.  There is
no
reason to think
that the notion of partial supersymmetry or our particular
approach to duality and its implications should not, in some fashion, be
extended to relativistic, interacting theories with a variety of spectra
(and
not simply those theories so closely tied to the Riemann zeta function
and other Dirichlet series), 
nor that we have exhausted the physical insights into number theories
that
consideration of quantum theories will provide.  We hope further work
will shed
light on these possibilities.

This research was supported in part by NSF Grant. No. PHY-9509991.

\listrefs
\bye